\documentclass[doublecol]{epl2}

\usepackage{graphicx}
\usepackage{dcolumn}
\usepackage{bm}
\usepackage{epsf}
\usepackage{amsmath}
\usepackage{amssymb}

\newcommand{\la}{\left\langle}
\newcommand{\ra}{\right\rangle}
\newcommand{\pd}{\partial}
\newcommand{\bla}{bla\\bla\\bla\\bla\\}

\newcommand{\PRL}{Phys. Rev. Lett. }

\title{Quantum fluctuation theorems in the strong damping limit}
\author{Sebastian Deffner, Michael Brunner and Eric Lutz}
\institute{Department of Physics, University of Augsburg, D-86135 Augsburg, Germany}

\abstract{
We consider a driven quantum particle in the strong friction regime described by the quantum Smoluchowski equation. We derive Crooks and  Jarzynski type relations  for  the reduced  quantum system by   properly generalizing the entropy production  to take into account the non-Gibbsian character of the equilibrium distribution. In the case of a nonequilibrium steady state, we obtain a quantum version of the Hatano-Sasa relation. We, further, propose an  experiment with driven Josephson junctions that would allow to investigate nonequilibrium entropy fluctuations in overdamped quantum systems.}

\begin{document}

\maketitle

Thermodynamic processes at the nanoscale are governed by both thermal  and quantum fluctuations. It has lately been recognized that for classical nanosystems the second law  of thermodynamics has to be generalized in order to include  effects induced by   thermal fluctuations. The latter are usually vanishingly small in macroscopic systems and are, therefore, neglected in the traditional formulation of thermodynamics \cite{bus05}. These generalizations of the second law take the form of fluctuation theorems that quantify the occurrence of negative fluctuations of  quantities like work, heat and entropy \cite{eva93,gal95}. A remarkable property of these new thermodynamic identities is their general validity arbitrarily far from equilibrium. An important example of a fluctuation theorem is the one derived by Crooks \cite{cro98}: it relates  the probability distributions of work, $\rho^F(W)$ and $\rho^R (W) $, along  \textit{forward} and  \textit{reversed} transformations of a system according to,
\begin{equation}
\label{1}
\rho^R (-W) = \rho^F (W)\exp{\left(-\beta( W -\Delta F)\right)}\ .
\end{equation} 
 Here $\Delta F$ is the free energy difference between  final and initial states. Equation \eqref{1} indicates  that large negative work fluctuations  are exponentially suppressed and, hence, not observable in macroscopic systems. In its integrated form, the  Crooks relation reduces to an equality previously obtained by Jarzynski \cite{jar97}, connecting the equilibrium free energy difference $\Delta F$ to the nonequilibrium work $W$ via, 
\begin{equation}
\label{2}
\left\langle \exp{\left(-\beta (W-\Delta F)\right)}\right\rangle=1 \ .
\end{equation}
In the above equation, the average $\la ... \ra$ is taken over the forward work distribution. It is essential to realize that Eqs.~\eqref{1} and \eqref{2} only apply for systems that are initially in an equilibrium Gibbs state. Extensions of these expressions for different nonequilibrium initial distributions have been introduced by Hatano and Sasa \cite{hat01} and by Seifert \cite{sei05}. Fluctuation theorems have been investigated experimentally in various nonequilibrium situations \cite{wan02,lip02,car04,tre04,col05,bli06}, where the canonical example consists  of a highly damped Brownian particle in a driven potential. Due to the experimental and theoretical importance of the strongly damped regime, the overdamped Langevin equation, and the equivalent Smoluchowski equation, have become the tool of choice for the analysis of classical fluctuation theorems.

In this paper, we derive quantum generalizations of the classical Crooks and Jarzynski relations, Eqs.~\eqref{1} and \eqref{2}, in the strong friction regime. Previous studies on isolated or weakly coupled quantum systems can be found in Refs.~\cite{muk03,mon05,def08,tal09}, while an extension to the strongly coupled regime has been recently put forward in Ref.~\cite{cam09}.  In the following,  we  use the quantum generalization of the Smoluchowski equation to treat both thermal and quantum fluctuations. Using the Onsager-Machlup path-integral representation, we show that the free energy difference for a driven quantum system can be obtained from its reduced semiclassical density operator. 
We, moreover, propose an experiment involving a driven Josephson junction that would allow to test our predictions.

\section{Quantum Smoluchowski equation} In the limit of high friction, the off-diagonal matrix elements $\la x|\hat \rho(t)|x'\ra$  of the system density operator in the position representation are strongly suppressed over a time scale of the order of $1/\gamma$, where $\gamma$ is the friction coefficient. As a result, a coarse-grained description of the  dynamics of the system in terms of the  diagonal part of the position distribution alone, $p(x,t) =  \la x|\hat \rho(t)|x\ra$,  becomes  possible \cite{ank01,cof07,mac04}. In this semiclassical picture, the quantum system follows a { classical} trajectory and quantum effects manifest themselves through quantum fluctuations that act in addition to  the thermal fluctuations induced by the heat bath.  A notable advantage of this description  is that the usual classical definitions of work and heat are valid,  in contrast to the full quantum regime \cite{tal07}. The quantum Smoluchowski equation can  be written as \cite{ank01,cof07,mac04},
\begin{equation}
\label{3}
\pd_t\,p(x,t)=\frac{1}{\gamma m}\pd_x\,\left[V'(x)+\frac{1}{\beta} \pd_x\,D_{e}(x) \right] p(x,t)\ ,
\end{equation}
where $V'(x)$ is   the derivative  of the external potential with respect to position and $\beta$ the inverse temperature.  The effective diffusion coefficient $D_e(x)$ is given by,
\begin{equation}
\label{4}
D_{e}(x)= 1/\left[1-\lambda \beta V''(x)\right]\,,
\end{equation}
with the parameter, 
\begin{equation}
\label{4a}
\lambda=\left(\hbar/\pi \gamma m\right)\left[c +\Psi\left(\gamma\hbar\beta/2\pi+1\right) \right]
\end{equation} 
measuring the magnitude of the quantum fluctuations. Here $m$ denotes the mass of the system, $c=0.577...$ is the Euler constant and $\Psi$  the digamma function \cite{abr65}. It should be noted that  quantum corrections depend explicitly on the position of the system through the curvature of the potential $V''(x)$. When $\lambda=0$, Eq.~\eqref{3} reduces to the classical Smoluchowski equation with constant diffusion coefficient.
The stationary equilibrium solution of Eq.~(\ref{3}), with natural boundary conditions, is
\begin{equation}
\label{5}
p_s(x)= \frac{1}{Z} \exp{\left(-\beta V(x)+\lambda\beta^2 V'(x)^2/2\right)}\left[1-\lambda \beta V''(x)\right]\,,
\end{equation}
where $Z$ is the normalization constant. The above equilibrium expression is in general non-Gibbsian when $\lambda \neq 0$.

The quantum Smoluchowski equation (\ref{3}) with the effective diffusion coefficient (\ref{4}) is valid in the semiclassical range of parameters,
\mbox{$\gamma/\omega_0^2\gg (\hbar\beta,1/\gamma)$}, \mbox{$\hbar\gamma\gg1/\beta$} and \mbox{$|\lambda\beta V''(x)|<1 $},
where $\omega_0$ is a characteristic frequency, i.e. curvature at a potential minimum  of the system \cite{ank01,cof07,mac04}. In the present analysis, we consider  a time-dependent problem where the  potential $V(x,\alpha_t)$ is driven by some external  parameter $\alpha_t= \alpha(t)$. The driving rate should be smaller than the relaxation rate, $\dot \alpha_t/\alpha_t \ll \gamma$, to ensure that  the non-diagonal elements of the density operator  remain negligibly small at all times \cite{dil09}. Note that this condition is not  restrictive in  the limit of very large $\gamma$.

\section{Quantum fluctuation theorems} We derive extensions of the Crooks and Jarzynski relations,  Eqs.~\eqref{1} and \eqref{2}, by using  a path integral representation of the solution of the quantum Smoluchowski equation  following Ref.~\cite{che06}. For the sake of generality, we  consider a generic driven Fokker-Planck equation, with  position-dependent drift and diffusion coefficients $D_1$ and $D_2$, of the form,
\begin{equation}
\label{7}
\pd_t\,p\left(x,\alpha,t\right)={ L}_\alpha \,p\left(x,\alpha,t\right)\ ,
\end{equation}
where the linear operator $L_\alpha$ is given by,
\begin{equation}
\label{8}
{L}_\alpha=-\pd_x\,D_1\left(x,\alpha\right)+\pd_x^2\,D_2\left(x,\alpha\right) \ .
\end{equation}
 The quantum Smoluchowski equation \eqref{3} corresponds to the  particular choice $D_1(x,\alpha) = -V'(x,\alpha)/\gamma m$ and $D_2(x,\alpha) = 1/[1-\lambda \beta V''(x,\alpha)]\gamma m\beta$. For any fixed value of the driving parameter $\alpha$, we write the stationary  solution of Eq.~\eqref{7} as,
\begin{equation}
\label{9}
p_s\left(x,\alpha\right)=\exp{\left(-\varphi\left(x,\alpha\right)\right)}\ ,
\end{equation}
where the function $\varphi(x,\alpha) $ is explicitly given by,
\begin{equation}
\label{10}
\varphi\left(x,\alpha\right)=\int^x dy \,\frac{\pd_y D_2\left(y,\alpha\right)-D_1\left(y,\alpha\right)}{D_2\left(y,\alpha\right)}\,. 
\end{equation}
We denote by $X=\{x\}^{+\tau}_{-\tau}$ a trajectory of the system that starts at $t=-\tau$ and ends at $t=+\tau$. We further define a \textit{forward process} $\alpha^F_t$, in which  the driving  parameter is varied from an  initial value  $\alpha^F_{-\tau}=\alpha_0$ to a final value $\alpha^F_{+\tau}=\alpha_1$, as well as its  time \textit{reversed process}, $\alpha^R_t=\alpha^F_{-t}$.
The conditional probability of observing a trajectory starting at $x_{-\tau}$ for the forward process can then be written as,
\begin{equation}
\label{11}
P^F\left[X|x_{-\tau}\right]=\exp{\left(-\int\limits_{- \tau}^{+\tau}dt\,S\left(x_t,\dot x_t, \alpha^F_t\right)\right)}\,,
\end{equation}
with a similar expression for the reversed process. In Eq.~\eqref{11} the generalized Onsager-Machlup function $S\left(x_t,\dot x_t, \alpha_t\right)$ is taken to be of the form \cite{risk},
\begin{equation}
\label{12}
S\left(x_t,\dot x_t, \alpha_t\right)=\frac{\left[\dot x_t-\left(D_1(x_t,\alpha_t)-\pd_x\,D_2(x_t,\alpha_t)\right) \right]^2}{4\, D_2(x_t,\alpha_t)}\ .
\end{equation}
The  last  term in the numerator of Eq.~\eqref{12} is included to guarantee that thermodynamic potentials are independent of the state representation \cite{gra79}, and follows from the It\^{o}-formula. By assuming that the system is initially in an equilibrium state given by the solution \eqref{9} of the Fokker-Planck equation \eqref{7},  we obtain that the net probability of observing the trajectory $X$ for the forward process is,
\begin{equation}
\label{13}
P^F[X]=p_s\left(x_{-\tau},\alpha_{0}\right)P^F\left[X|x_{-\tau}\right]\,.
\end{equation}
In complete analogy, we find that the  corresponding unconditional probability for the reversed process reads,
\begin{equation}
\label{14}
P^R[X]=p_s\left(x_{\tau},\alpha_1\right)P^R\left[X^\dagger|x_{\tau}\right]\,,
\end{equation}
where we have introduced the time-reversed trajectory, \mbox{$X^\dagger =\{x^\dagger_t\}^{+\tau}_{-\tau}$} with $x^\dagger_t=x_{-t}$. We next  compare the probability of having the trajectory $X$ during the forward process with that of having the trajectory $X^\dagger$ during the reversed process. We have 
\begin{equation}
\label{15}
\begin{split}
P^R \left[X^\dagger|x^\dagger_{-\tau}\right]&=\exp{\left(-\int\limits_{- \tau}^{+\tau}dt\,S\left(x_t^\dagger,\dot x_t^\dagger, \alpha_t^R\right)\right)}\\ 
&=\exp{\left(-\int\limits_{- \tau}^{+\tau}dt\,S^\dagger\left(x_t,\dot x_t, \alpha_t^F\right)\right)}\,, 
\end{split}
\end{equation}
where we have defined the conjugate Onsager-Machlup function, \mbox{$S^\dagger\left(x_t,\dot x_t, \alpha_t\right)=S\left(x_t,-\dot x_t, \alpha_t\right)$}. The ratio of the conditional probabilities (\ref{11}) and (\ref{15}) is simply determined by  the difference of $S$ and $S^\dagger$. Using the definition (\ref{12}), we thus obtain,
\begin{equation}
\label{16}
\begin{split}
\frac{P^F\left[X|x_{-\tau}\right]}{P^R \left[X^\dagger|x^\dagger_{-\tau}\right]}&=\exp{\left(\int\limits^{+\tau}_{-\tau} dt\,\frac{D_1 \left(x_t,\alpha^F_t\right)}{D_2 \left(x_t,\alpha_T^F\right)} \dot x_t)\right)}\\
&\times \exp{\left(-\int\limits^{+\tau}_{-\tau} dt\,\frac{\pd_x\,D_2 \left(x_t,\alpha^F_t\right) }{D_2 \left(x_t,\alpha^F_t\right)} \dot x_t\right) }\ .
\end{split}
\end{equation}
The ratio of  the forward and reversed probabilities, Eqs.~\eqref{13} and \eqref{14}, follows directly as,
\begin{equation}
\label{17}
\begin{split}
\frac{P^F\left[X\right]}{P^R \left[X^\dagger\right]}&=\frac{p_s\left(x_{-\tau},\alpha_0 \right)P^F\left[X|x_{-\tau}\right]}{p_s\left(x_{\tau},\alpha_1\right)P^R \left[X^\dagger|x^\dagger_{-\tau}\right]}\\
&=\exp{\left(\Delta \varphi +\int\limits^{+\tau}_{-\tau} dt\,\frac{D_1 \left(x_t,\alpha^F_t\right)}{D_2 \left(x_t,\alpha^F_t\right)}\, \dot x_t\right)}\\
&\times\exp{\left(-\int\limits^{+\tau}_{-\tau} dt\,\frac{\pd_x\,D_2 \left(x_t,\alpha^F_t\right) }{D_2 \left(x_t,\alpha^F_t\right)} \dot x_t\right)}\, ,
\end{split}
\end{equation}
where 
\begin{equation}
\label{18}
\Delta \varphi=\int\limits^{+\tau}_{-\tau} dt\left(\dot\alpha_t^F\,\pd_\alpha\varphi+\dot x_t\, \pd_x\varphi\right)\, .
\end{equation}
By using the explicit expression \eqref{10} of the stationary solution $\varphi(x,\alpha)$, we finally arrive at
\begin{equation}
\label{19}
\frac{P^F\left[X\right]}{P^R \left[X^\dagger\right]}=\exp{\left(\int\limits^{+\tau}_{-\tau} dt\,\dot\alpha_t^F\,\pd_\alpha\varphi\right)}\,.
\end{equation}
We are now in the position to derive generalized fluctuation theorems for  stochastic processes described by the generic Fokker-Planck equation \eqref{7}. We begin by defining  the generalized entropy production $\Sigma$ as,
\begin{equation}
\label{20}
\Sigma =\int\limits_{-\tau}^\tau dt\,\dot\alpha^F_t \, \pd_\alpha\varphi \, .
\end{equation}
The  entropy production $\Sigma$ in Eq.~(\ref{20}) is similar to the entropy production introduced by Hatano and Sasa for systems initially in a nonequilibrium steady state \cite{hat01}. In the present situation, however, it corresponds to a non-Gibbsian equilibrium state. We note, in addition, that the entropy production,  as defined in Eq.~(\ref{20}), is odd under time-reversal, \mbox{$\Sigma^R\left[X^\dagger\right]=-\Sigma^F\left[X\right]$}. The  distribution of the  entropy production, $\rho^{F}(\Sigma)$, for an ensemble of realizations of forward  processes can then be defined as,
\begin{multline}
\label{21}
\rho^F\left(\Sigma\right)=\int {\cal D}X\,P^F\left[X\right]\delta\left(\Sigma-\Sigma^F\left[X\right]\right) =\\
=\exp{\left(\Sigma\right)}\int {\cal D}X^\dagger\,P^R\left[X^\dagger\right]\delta\left(\Sigma+\Sigma^R\left[X^\dagger\right]\right) 
\end{multline}
where we have used  Eq.~\eqref{19} in the last line. Here, $\int {\cal D}X=\lim\limits_{N \rightarrow \infty}{\left(4 \pi s\right)^{-N/2}\prod\limits_{i=1}^{N-1}\int dx_{is} D\left(x_{is},\alpha_{is}\right)^{-1/2}}$, $s=2 \tau/N$, denotes  the product of integrals over all possible paths $X$. The continuous integral in Eq.~\eqref{21} is  interpreted as the limit of a discrete sum. Equation \eqref{21} can be recast in the form of a generalized Crooks relation for the entropy production, 
\begin{equation}
\label{22}
\rho^R\left(-\Sigma\right) = \rho^F\left(\Sigma\right)\exp{\left(-\Sigma\right)}\,.
\end{equation}
By, moreover, integrating Eq.~(\ref{22}) over $\Sigma$, we obtain an extended version of the Jarzynski equality,
\begin{equation}
\label{23}
\left\langle \exp{\left(-\Sigma\right)} \right\rangle=1\ .
\end{equation}
Expression \eqref{20} for the entropy production, together with the  fluctuation theorems \eqref{22} and \eqref{23}, constitutes  our main result.  Combined, they represent the quantum  generalizations of the Crooks and Jarzynski equalities, Eqs.~\eqref{1} and \eqref{2},  in the limit of strong damping.   In the classical limit $\lambda = 0$, $\varphi(x,\alpha) = \beta (V(x,\alpha) -F(\alpha))$, and  the entropy production  \eqref{20} takes the familiar form, $\Sigma=\beta \int dt\,\dot\alpha^F_t\, \pd_\alpha V(x_t,\alpha^F_t) - \beta\,\Delta F$.   The inequality $\la \Sigma \ra \geq 0$ implied by Eq.~\eqref{23} is often  interpreted as an expression of the second law. It is worthwhile to mention that the above derivation applies without modification to the case of   an initial nonequilibrium steady state, instead of an initial equilibrium state, leading directly  to a quantum generalization of the Hatano-Sasa relation \cite{hat01}.

\section{Parametric harmonic oscillator}
Let us illustrate our results with the example  of a harmonic oscillator with time-dependent frequency, $ V(x,\omega_t) = m\omega_t^2 x^2/2$. The effective diffusion coefficient \eqref{4} is  in this case position independent,  $D_{e}(x)= 1/(1-\lambda \beta m \omega^2)$,  and quantum fluctuations therefore  renormalize the width of the stationary distribution  of the oscillator, which is no longer given by the temperature of  the bath as in the classical regime. We assume that  the  driving parameter $\alpha_t = \omega_t^2$ is changed from $\omega_0^2$ to $\omega_1^2$ during time $2\tau$. We moreover  define  the partition function $Z$ of the system as the normalization constant of the stationary distribution \eqref{5} of the quantum Smoluchowski equation. The free energy difference between final and initial state is then,
\begin{equation}
\beta\Delta F = -\ln (Z_1/Z_0)= \ln\frac{\omega_1}{\omega_0} + \ln \sqrt{ \frac{1-\lambda\beta m \omega_0^2}{1-\lambda\beta m \omega_1^2} } \ .
\label{24}
\end{equation}
In the limit $\hbar \gamma\beta \gg 1$, the quantum parameter \eqref{4a} simplifies to  $\lambda =(\hbar/\pi  \gamma m ) [\ln( \gamma/\nu)+c]$ with $\nu = 2\pi/\hbar\beta$.  The free energy difference \eqref{24} reduces accordingly to
\begin{equation}
\beta\Delta F \simeq \ln\frac{\omega_1}{\omega_0} + \frac{\ln(\gamma/\nu)}{\nu\gamma} \left( \omega_1^2 - \omega_0^2 \right) 
+ \frac{c}{\nu\gamma} \left( \omega_1^2 - \omega_0^2 \right) \ .
\label{25}
\end{equation}
The generalized Jarzynski equality  \eqref{23} can now be used to evaluate the free energy difference  for the parametric quantum oscillator from nonequilibrium work measurements.  For simplicity, we consider a linear variation of the square frequency, $\omega_t^2 = \omega_0^2+ (\omega_1^2-\omega_0^2)t/2\tau$. We define the semiclassical work as  $\beta w= \int_0^{2\tau}dt\,\dot\alpha_t \, \pd_\alpha\tilde\varphi$ with $\tilde\varphi =\beta V(x, \alpha) -(\lambda \beta^2/2) V'(x,\alpha)^2 -\ln(1-\lambda \beta V''(x,\alpha))$; the  latter is related to the entropy production \eqref{20} via $\beta w= \Sigma +\beta \Delta F$ and reduces to the classical expression $\beta W$ of the work in the limit $\lambda=0$. We numerically determine the probability distribution $\rho(\beta w)$ of the work   from an ensemble of identical driving realizations with  the help of the It\^o-Langevin equation, $m \gamma \dot x +V'(x,\alpha_t)= \sqrt{2 D_e(x,t)} F(t)$, corresponding to the quantum Smoluchowski equation \eqref{3}. Here $F(t)$ denotes a Gaussian random force with zero mean and variance $\la F(t)F(t')\ra = m\gamma/\beta\, \delta(t-t')$.  
\begin{figure}
\includegraphics[width=0.49\textwidth,angle=0]{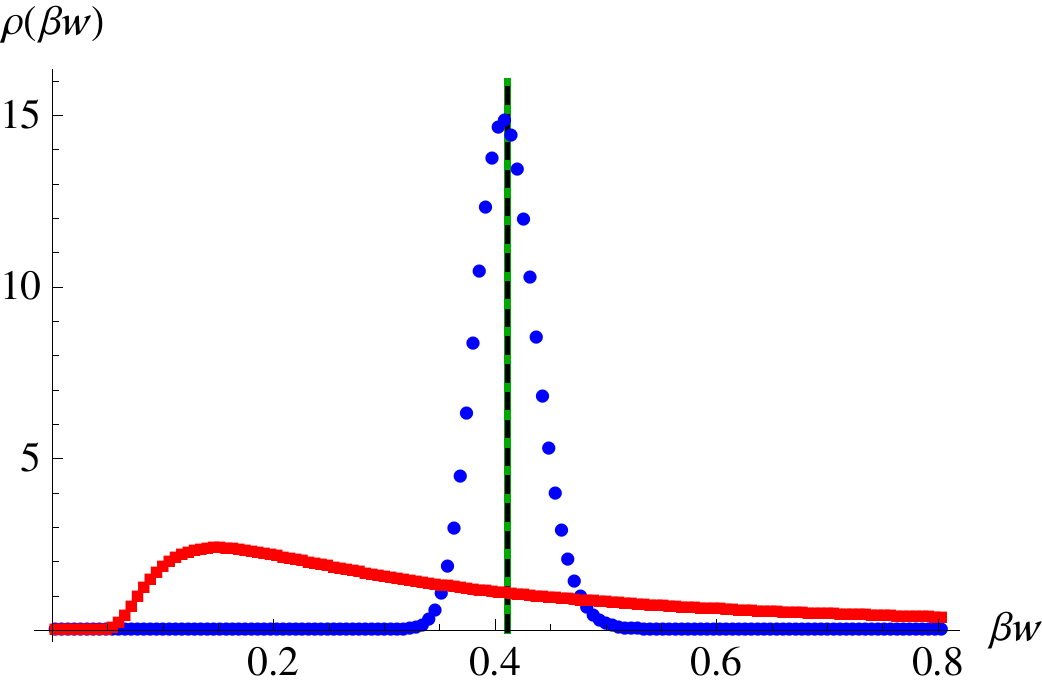}
\caption{(color online)\label{f1} Work distribution $\rho(\beta w)$ for a quantum oscillator with time-dependent  frequency, $\omega_t^2 = \omega_0^2+ (\omega_1^2-\omega_0^2)t/2\tau$, for  slow driving, $2\tau = 300\tau_r$ (blue dots),  and  fast driving, $2\tau = \tau_r$ (red squares), with $\tau_r =1/\gamma$  the relaxation time of  the oscillator. In both cases, the free energy difference evaluated numerically using   the Jarzynski equality \eqref{23} (green, solid vertical line) agrees with the analytical expression \eqref{24} (black, dashed vertical line). Parameters are $\hbar=1.05 $, $m=1$,   $\gamma=6000$, $\beta = 103.46$, $\omega_0^2=5$  and $\omega_1^2=11$, for an ensemble of $2 \cdot 10^7$ trajectories.}
\end{figure}
Figure \ref{f1} shows the work distribution $\rho(\beta w)$ for two different driving times: a slow driving ($2\tau = 300\tau_r$)  and  a fast driving  ($2\tau = \tau_r$), where $\tau_r = 1/\gamma$ is the relaxation time of  the oscillator. We observe that   equality \eqref{23} leads in both cases to the  free energy difference \eqref{24}, whose value is indicated by the vertical line.  It is worth noticing that a naive application of the classical Jarzynski equality \eqref{2} to the quantum oscillator would result in an apparent violation of the latter \cite{che04}; this deviation is of course due to the non-Gibbsian property of the  stationary distribution \eqref{5}.
 
In the approach of Ref.~\cite{cam09}, the free energy $F_S$ of the system is defined as the difference between the total free energy of system plus bath and the  free energy of the bath alone. The corresponding free energy difference can be evaluated exactly and reads \cite{gra84,ing02},
\begin{equation}
\beta\Delta F_S = \ln \frac{\omega_0 \Gamma \left( \frac{\lambda_1(\omega_0)}{\nu} \right) \Gamma \left( \frac{\lambda_2(\omega_0)}{\nu} \right) }{\omega_1 \Gamma \left( \frac{\lambda_1(\omega_1)}{\nu} \right) \Gamma \left( \frac{\lambda_2(\omega_1)}{\nu} \right) } \ ,
\label{25a}
\end{equation}
where $\Gamma(x)$ is the Gamma function and $\lambda_{1,2}$ the characteristic frequencies of the damped oscillator; in the limit of large bath cutoff frequency, they are given by  $\lambda_{1,2}(\omega) = \gamma/2 \pm \sqrt{\gamma^2/4 - \omega^2}$ \cite{gra84}.  Using the asymptotic expansions of the Gamma function, $\Gamma(x) \simeq 1/(x + c x^2) (x\ll   1)$ and $\Gamma(x) \simeq \sqrt{2\pi}\, x^{x-1/2}\exp(-x) (x \gg  1)$ \cite{abr65},  expression \eqref{25a} is seen to reduce to the free energy difference \eqref{25} obtained from the quantum Smoluchowski equation in the limit $\gamma\gg \hbar\beta\omega_i^2$ and  $\gamma\gg \omega_i$. 
\begin{figure}
\includegraphics[width=0.49\textwidth,angle=0]{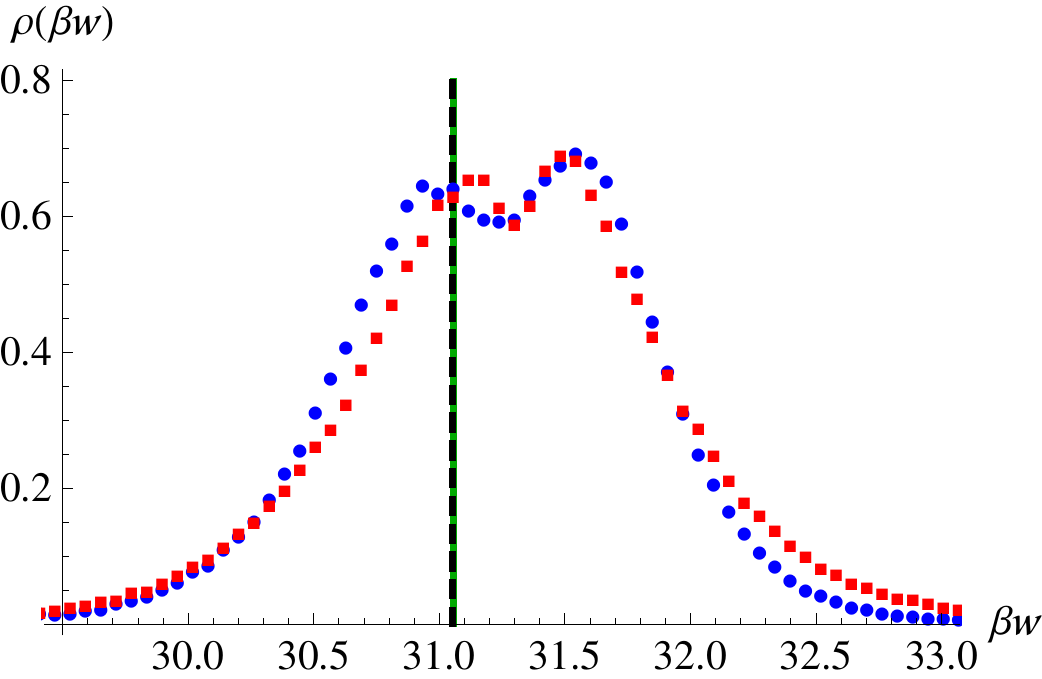}
\caption{\label{f2}(color online) Work distribution $\rho(\beta w)$ for a driven Josephson junction with potential $V(\phi,\alpha_t) =-E_J (\cos(\phi) + \alpha_t \phi)$ and  driving parameter $\alpha_t = \alpha_0+(\alpha_1-\alpha_0)t/2\tau$, for fast driving, $2 \tau  = 0.2 \gamma/E_J$ (red squares), and  slow driving, $2\tau  =  \gamma/E_J$ (blue dots). The free energy difference evaluated numerically using the
Jarzynski equality (23) (green, solid vertical line) agrees with the one
determined via the normalization constant of the stationary
distribution (30) (black, dashed vertical line). Parameters are $\hbar = 1.05$, $m = 1$, $\gamma= 22.5$,   $\beta = 0.72$,
 $\alpha_0 = 0$ and $\alpha_1 = 0.3$, $E_J = 50.64$, for an ensemble of $2 \cdot   10^5$
trajectories. The latter correspond to $T=1$, $\lambda=0.026$ and $\Theta=0.96$.}
\end{figure}

\section{Experimental verification in driven Josephson junctions} 
No experimental investigation of quantum fluctuation theorems has been performed so far. A scheme to study  the Crooks and Jarzynski relations in isolated and weakly damped quantum systems using modulated ion traps has  recently been  put forward  in Ref. \cite{hub08}. Here we  propose to test the predictions for the quantum fluctuation theorems, Eqs.~\eqref{22} and  \eqref{23}, in the strong damping limit using driven  Josephson junctions \cite{lik}. 
The  Josephson relations for the current $I_s(t)$ across the junction and the phase difference $\phi(t)$ between left and right superconductors are 
\begin{equation}
\label{26}
I_s=I_c\sin{\left(\phi\right)}\hspace{.6em} \text{and}\hspace{.7em} \dot\phi=2 e/\hbar\,\,U(t)  \ ,
\end{equation}
where $U(t)$ is the voltage drop across the junction. The maximal current $I_c$ is given by \mbox{$I_c=\left(2e/\hbar \right) E_J $}, where $E_J$ is the coupling energy (Josephson energy). An externally shunted Josephson junction can be described via an equivalent circuit consisting of an ideal junction, a capacitance $C$ and a resistance $R$ (Resistively Shunted Junction (RSJ) model) \cite{lik}. In this model, the Josephson junction is interpreted as describing the diffusive motion of a particle with position $\phi(t)$ and mass $m=(\hbar/2e)^2 C$, the  friction coefficient being given by $\gamma=1/RC $. An important quantity in the RSJ model is the dimensionless capacitance (Stewart-McCumber parameter), $\beta_c=(2\pi/\Phi_0)\,I_c R^2 C$, where $\Phi_0=h/2e$ is the magnetic flux quantum. In the overdamped regime,  $\beta_c<1$, the dynamics of the Josephson phase $\phi$  can be described by the quantum Smoluchowski equation \eqref{3} with the potential $V(\phi)=-E_J\, \cos{(\phi)}-E_I\, \phi$  \cite{ank03,ank05}. The energy $E_I=(\hbar/2 e) I$ is determined by the bias current $I$    and  the effective diffusion coefficient \eqref{4} reads,
\begin{equation}
\label{29}
D_e(\phi)=1/[1-\Theta \cos{(\phi)}]\ .
\end{equation}
The constant $\Theta=\lambda \beta E_J$ is the crucial parameter which governs the magnitude of quantum effects in a Josephson junction. It is directly proportional to the quantum parameter $\lambda$, Eq.~\eqref{4a}, which  in the context of the RSJ model can be reexpressed as,
\begin{equation}
\lambda=2 r \left[c+\Psi\left(\beta E_c/2 \pi^2 r +1\right) \right]\,,
\end{equation}
where \mbox{$E_c=2 e^2/C$} is the charging energy, $r=R/R_Q$ the dimensionless resistance and   \mbox{$R_Q=h/4 e^2$} the resistance quantum. The stationary solution of the quantum Smoluchowski equation \eqref{3}, with periodic boundary conditions, with $V'(\phi) = V'(\phi+L)$, can be written as \cite{rei02},
\begin{equation}
\label{30}
p_{stat}(\phi) = \frac{p_s(\phi) }{Z_J} \int_\phi^{\phi+L} \!dy \,[D_e(y)p_s(y)]^{-1}=\frac{1}{Z_J} e^{-\tilde\varphi} \ , 
\end{equation}
where $p_s(\phi) $ is given by Eq.~\eqref{5} and $Z_J$ is the normalization constant.
In Fig.~\ref{f2} we have plotted the work distribution $\rho(\beta w)$ for a driven Josephson junction with potential  $V(\phi,\alpha_t) =-E_J (\cos(\phi) + \alpha_t \phi)$ and linear driving parameter $\alpha_t = E_I(t)/E_J = \alpha_0+(\alpha_1-\alpha_0)t/2\tau$, for  two driving times, $2\tau = 0.2 \gamma/E_J$ and  $2\tau = \gamma/E_J$. As in the case of the parametric quantum oscillator, the free energy difference determined numerically using the generalized Jarzynski equality \eqref{23} agrees with the one determined via the normalization constant of the stationary distribution \eqref{30}, $\beta \Delta F = -\ln (Z_{J,1}/Z_{J,0})$, of the quantum Smoluchowski equation. 
\begin{table}
\begin{center}
\begin{tabular}{c|c|c|c}
&$T[K]$ & $\lambda[10^{-3}]$&$\Theta$  \\
\hline
Quantum&0.98 & 0.21 &0.99\\
Classical& 4.2 &0.087 &0.097\\
\end{tabular}
\caption{\label{tab1}Typical $\Theta$ values for circle shaped Josephson junctions with $C=1.2 pF$, $R=0.37 \Omega$, $I_c=0.2 mA$ and $\beta_c=0.1$.}
\end{center}
\end{table}

The nonequilibrium entropy production \eqref{20} can be experimentally determined in a Josephson junction by applying the following measurement procedure. The  phase $\phi$ can be directly deduced from a measurement of the Josephson current once the current--phase relation of the junction has been determined \cite{gol04}. The system is then first prepared in a given initial state and let to relax to its stationary state \eqref{5}. After the latter has been attained, the Josephson potential $V(\phi)$ is modified according to a specific driving protocol $\alpha_t$ with the help of an external magnetic field. The entropy production $\Sigma$ during such a protocol (corresponding to either a {\it forward} or {\it reversed} transformation) can  be evaluated via Eq.~\eqref{20} from the recorded values of the current.   The distribution function of the entropy can eventually be reconstructed by repeating the above measurement sequence, and the validity of the quantum fluctuation theorems \eqref{22} and \eqref{23} in the strong friction regime can be tested. 
In Tab.~\ref{tab1} we list typical parameter values for niobium-based Josephson junctions \cite{kai10}. By varying the temperature, both the  classical, $\Theta \ll 1$, and the  quantum regime, $\Theta \lesssim 1$, can be explored with the same junctions.

\section{Conclusion} 
We have analyzed quantum fluctuation theorems in the strong coupling limit with the help of the quantum Smoluchowski equation. We have shown that quantum  Crooks and Jarzynski type relations can be derived  in this regime  when the entropy production is properly modified to take into account the non-Gibbsian property of the initial equilibrium state.  In the case of an initial nonequilibrium steady state, a similar calculation leads to a quantum Hatano-Sasa relation. By investigating a parametric harmonic oscillator and a driven Josephson junction, we have additionally  shown that  the free energy difference can be directly obtained from the reduced density operator of the quantum system.
We have, finally, proposed an  experiment based on a  driven Josephson junction that would enable to study quantum nonequilibrium entropy fluctuations in overdamped systems.

\acknowledgments{
We thank P. Talkner for bringing Ref.~\cite{gra79} to our attention,  C. Schneider and R. Held, as well as  C. Kaiser and R. Sch\"afer for their experimental advice on Josephson junctions.  This work was supported by  the Emmy Noether Program of the DFG (contract No LU1382/1-1) and the cluster of excellence Nanosystems Initiative Munich (NIM).}

\end{document}